\documentclass[12pt]{article}
\usepackage{graphics}
\usepackage{amssymb}
\usepackage{color}
\usepackage{amsmath}
\usepackage{amsthm}
\usepackage{amsopn}
\def\uno{\mbox{1 \kern-.59em {\rm l}}}

\def\be{\begin{equation}}
\def\ee{\end{equation}}
\def\bea{\begin{eqnarray}}
\def\eea{\end{eqnarray}}

\begin{document}
\begin{center}
{\bf{\large Neutral current neutrino oscillation  via quantum field theory approach}}
\vskip 4em
{\bf M. M. Ettefaghi} \footnote{ mettefaghi@qom.ac.ir  }
 \vskip 1em
 Department of Physics, University of Qom, Qom 371614-6611,
Iran.
\vskip 4em
{\bf Z. Askaripour Ravari}
\vskip 1em
University of Kashan, Km 6 Ravand Road, Kashan 87317-51167, IRAN
 \end{center}
 \vspace*{1.9cm}
\begin{abstract}
Neutrino and anti-neutrino states coming from the neutral current or $Z_0$ decay are blind with respect to the flavor. The neutrino oscillation is observed and formulated when its flavor is known. However, it has been shown that we can see neutrino oscillation pattern for $Z_0$ decay neutrinos provided that both neutrino and anti-neutrino are detected. In this paper, we restudy this oscillation via quantum field theory approach.  Through this approach, we find that the oscillation pattern ceases if  the distance between the detectors is larger than the coherence length, while both neutrino and antineutrino states may be coherent. Also the uncertainty of source (region of $Z_0$ decay) does not have any role in the coherency of neutrino and antineutrino.
\end{abstract}
\section{Introduction}
 While neutrino oscillation is a window to the new physics, it is one of the most interesting quantum mechanical phenomena.
  Historically, neutrino oscillation has been established more than 50 years and confirmed experimentally more than 10 years \cite{historical}.
  The observation of neutrino oscillation depends on the coherency of neutrinos during the production, propagation and detection \cite{paradoxes,coherence}. The production and detection coherence conditions are satisfied provided that the intrinsic quantum
mechanical energy uncertainties during these processes are large compared
to the energy difference $\Delta E_{jk}$ of different neutrino mass eigenstates:
\be
\Delta E_{jk}\sim\frac{\Delta m^2_{jk}}{2E}\ll\sigma_E,
\ee
where $\sigma_E=\mbox{min}\{\sigma^{prod}_E,\sigma^{det}_E\}$. This condition implies that, during the production and detection processes, one cannot discriminate the neutrino mass eigenstates.
Conservation of the coherency during the propagation means that the wave packets describing the mass eigenstates overlap from the production until the detection regions. The wave packets describing the different neutrino mass eigenstates propagate with different group velocities.
 After propagating $L$, the separation of different mass wave packets is $\frac{\Delta m_{ij}^2}{2E^2}L$.  Consequently, the coherent propagation is guaranteed provided that
\be
\frac{\Delta m^2_{ij}}{2E^2}L\ll\sigma_{x\nu}\simeq\frac{v_g}{\sigma_E},
\ee
where $v_g$ is the average group velocity of the wave packets of different
neutrino mass eigenstates and $\sigma_{x\nu}$ is their common effective spatial width.
In other words, similar to the double-slit experiment, if one could determine which mass eigenstate is created or detected, the neutrino oscillation pattern would disappear. For instance, conservation of energy and momentum implies that exact determination of energy-momentum of charged leptons leads to the determination of mass eigenstate of the corresponding neutrino (in fact exact momentum conservation causes neutrino state is entangled kinematically to the corresponding charged lepton state) and neutrino oscillation is ceased \cite{Found}.

The kinematics analysis shows that a neutrino state created through charged current interactions has a specific flavor. For instance, muon neutrino is created by the pion decay while muon decay gives only electron neutrino.
 In contrast, neutral current or
$Z^0$ decay is blind with respect to the neutrino flavors. In other words, every flavor eigenstate as well as every mass eigenstate is created with the same probability. However, there is another property that is noticeable; neutrino and antineutrino states are entirely correlated in the sense that they have same flavor. It has been shown that if both neutrino and antineutrino are detected, one can observe neutrino oscillation pattern between the detectors \cite{smirnov}. Nevertheless, if only either neutrino or antineutrino is detected, the neutrino oscillation is ceased; therefore, it is a realization of the Einstein-Podolsky-Rosen paradox \cite{epr}.
In the center of mass frame, in particular, the oscillation pattern occurs at distance $L+\bar{L}$, where $L$ and $\bar{L}$ are the distance of the neutrino and antineutrino detectors from the source, respectively.
In this paper, we reanalyze it via quantum field theory approaches.
In this approach, the oscillating states become intermediate states, not directly observed, which propagate between a source and a detector. The localization conditions are respected with attributing a localized wave function to interacting initial and final states in the source and detector \cite{CC,16,17,Beuthe,qmqft}. Indeed, these localizations are essential for the observation of the neutrino oscillation and guarantee the coherence issues \cite{Found}. In the case of $Z_0$ decay neutrinos, we will see the localization of source (region of $Z_0$ decay) is not important while the localization of detectors plays role in the coherency condition of neutrino and antineutrino.
Moreover, as was said in general, the coherency is spoiled during propagation because the group velocities of various mass eigenstates are different. Therefore, maybe one expects that when both neutrino and antineutrino propagate coherently, it is possible to have oscillation pattern. However, we will show that it is necessary the distance between detectors to be smaller than the coherence length.

In the following, we develop $Z_0$ decay neutrino oscillation through quantum field theory approach. Finally, we discuss on the coherency properties which appears through quantum field theory approach.

\section{Developing neutral current neutrino oscillation through quantum field theory approach}\label{ft}
We can describe any particle physics processes by S-matrix formalism in quantum field theory provided that it is adjusted according to the physical situations. In particular, to describe neutrino oscillation one needs to notice that neutrinos are produced and detected in confined space-time regions. The source and detector regions are separated by a finite distance which is usually much larger than the size of these regions.

Neutral current neutrino oscillation consists of the following three processes:
\begin{itemize}
\item Creation neutrino and antineutrino in the source $$Z_0\rightarrow \nu +\bar \nu.$$
\item Detection of neutrino in the corresponding detector $$\nu + D_I\rightarrow D_F+l^-.$$
\item Detection of antineutrino in the other detector $$\bar \nu + \bar{D}_I \rightarrow \bar{D}_F + l^+.$$
\end{itemize}
 In order to define the initial and final states, the localizations of interactions in the source and detectors require to integrate on momentum with a localized distribution function around the corresponding averaged momentum.  Therefore, the initial states are defined as follows:
\bea
\vert{Z_0}\rangle =\int[d\mathbf{p}] F_z(\mathbf p,\mathbf P)\vert {Z_0,\mathbf p}\rangle \ ,\nonumber\\
\vert {D_I}\rangle =\int[d\mathbf k]F_{D_I}(\mathbf k,\mathbf K)\vert {D_I,\mathbf k}\rangle \ ,\nonumber\\
\vert {\bar D_I}\rangle =\int[d \mathbf {\bar {k}}]F_{\bar D_I}(\mathbf {\bar {k}},\mathbf{\bar{ K}})\vert {\bar D_I, \mathbf {\bar{ k}}}\rangle \ ,\label{iket}
\eea
where $D_I$ ($\bar D_I$) is target in the detectors of neutrino (antineutrino). $[d\mathbf p]$ denotes $\frac{d^3\mathbf p}{(2\pi)^3\sqrt{2E_p} }$.
The final states are written similarly as follows:
\bea\label{fket}
\vert {D_F}\rangle =\int[d\mathbf k'] F_{D_F}(\mathbf k',\mathbf K')\vert {D_F,\mathbf k'}\rangle \ ,\nonumber\\
\vert {\bar D_F}\rangle =\int[d \mathbf{\bar{k'}}]F_{\bar D_F}(\mathbf{\bar{ k'}},\mathbf{\bar{ K'}})\vert {\bar D_F, \mathbf{\bar {k'}}}\rangle \ ,\nonumber\\
\vert {l^-}\rangle =\int[d\mathbf k''] F_{l^-}(\mathbf k'',\mathbf K'')\vert {l^-,\mathbf k''}\rangle \ ,\nonumber\\
\vert {l^+}\rangle =\int[d \mathbf{\bar{k''}}]F_{l^+}(\mathbf{\bar{ k''}},\mathbf{\bar{ K''}})\vert {l^+, \mathbf{\bar {k''}}}\rangle \ .
\eea
Here, $D_F$ ($\bar D_F$) refers to created nucleon (antinucleon) in detector due to neutrino (antineutrino) collision. $l^-$ ($l^+$) denotes the created charged lepton corresponding to neutrino (antineutrinos) in the detector. In the above defined states, $F$'s are momentum distribution functions which are localized around the corresponding mean momentum. The amplitude of the neutrino and anti-neutrino production - propagation - detection processes is given by the following matrix element:
\bea\nonumber
i\mathcal A_{\alpha\beta}&\!\! =\!\! &\langle {D_F,l^-,\bar D_F,l^+}\vert {\widehat T e^{-i\int{d^4x \mathcal H_I(x)}}- 1} \vert {D_I,\bar D_I,Z_0}\rangle \nonumber\\
&\!\! =\!\! &\langle {D_F,l^-,\bar D_F,l^+}\vert {i \int d^4x_1\int d^4x \int d^4x_2 \mathcal H_I(x_2)\mathcal H_I(x)\mathcal H_I(x_1)} \vert {D_I,\bar D_I,Z_0}\rangle \nonumber\\
&\!\! =\!\! & \dfrac{1}{\sqrt{3}}\sum _j U_{\alpha j}^*U_{\beta j}\langle {D_F,l^-,\bar D_F,l^+}\vert {i\mathcal A_j^{p.w.}}\vert {D_I,\bar D_I,Z_0}\rangle  \ .
\label{amp}
\eea
where $\hat T$ is the time ordering operator and $\mathcal H_I$ are the weak interaction Hamiltonian. The quantity $\mathcal A_j^{p.w.}$ is the plane wave amplitude of the process with the $j$'th neutrino and antineutrino mass eigenstate propagating between the source and the detectors and is written as follows:
\bea\nonumber
i\mathcal A_j^{p.w.}(k,k',k'',\bar k, \bar k', \bar k'') &\!\! =\!\! & i\int d^4x_1\int d^4x \int d^4x_2 \widetilde M_{j\bar D}(\bar k,\bar k', \bar k'') e^{-i(\bar k-\bar k'- \bar k'')x_2}\nonumber\\
&\!\! \times \!\! &\int \dfrac{d^4\bar q}{(2\pi)^4}\dfrac{\bar q\!\!\!/ + m_j}{\bar q^2-m^2_j+i\epsilon} e^{-i\bar q(x_2-x)}
\widetilde M_{jjZ}( p) e^{-ipx}\nonumber\\
&\!\! \times \!\! &\int {\dfrac{d^4q}{(2\pi)^4}\dfrac{q\!\!\!/ +m_j}{q^2-m_j^2+i\epsilon}e^{-iq(x_1-x)}}\nonumber\\
&\!\! \times \!\! &\widetilde M_{jD}( k, k',k'') e^{-i(k-k'-k'')x_1} \ ,
\eea
 where $\widetilde M$'s are the plane wave amplitudes of the processes. It is convenient to switch to shifted 4-coordinate variables $x$, $x_1$ and $x_2$ defined according to
$$x_1\rightarrow x_1+x_D~~~~~~x_2\rightarrow x_2+x_{\bar{D}} ~~~~~~x\rightarrow x+x_P,$$
where the propagation times $T$ and $\bar T$ are defined by
$$ t_D-t_P=T,~~~~~~~~~~~t_{\bar{D}}-t_P=\bar T,$$
and the propagation distances by
$$\mathbf{x}_D-\mathbf{x}_P=\mathbf{L},~~~~~~~~~~~\mathbf{x}_{\bar{D}}-\mathbf{x}_P=\mathbf{\bar L}, $$
and we redefine
$$
F(k)=f(k)e^{ikx_D}\Longrightarrow f(k)=F(k)e^{-ikx_D}.
$$
Taking into account that
$$
q\!\!\!/ +m_j=\sum u_j(q,s)\bar u_j(q,s),
~~~~~~~~~
\bar {q\!\!\!/}  -\bar m_j=\sum v_j(\bar q,s)\bar v_j(\bar q,s),
$$
we redefine the amplitudes (including spinors) correspond to the production and neutrino and antineutrino detection processes, respectively, as follows:
$$M_{jjZ}(p)=\dfrac{\bar v_j(\bar q)}{\sqrt{2\bar {q}_0}}\widetilde M_Z( p)\dfrac{u_j(q)}{\sqrt{2q_0}},$$
$$M_{jD}(k,k',k'')=\dfrac{\bar u_j(q)}{\sqrt{2q_0}}\widetilde M_D(k,k',k''),$$
$$M_{j\bar D}(\bar k,\bar k', \bar k'')=\widetilde M_{\bar D}(\bar k,\bar k', \bar k'')\dfrac{v_j(\bar q)}{\sqrt{2\bar q_0}}.$$
Substituting the initial and final state from (\ref{iket}) and (\ref{fket}) into (\ref{amp}) and using above issues, we have
\bea\nonumber
\mathcal A_{\alpha\beta} &\!\! =\!\! & \dfrac{1}{\sqrt{3}}\sum _j U_{\alpha j}^*U_{\beta j}\int \dfrac{d^4q}{(2\pi)^4}\dfrac{2q_0}{q^2-m_j^2+i\epsilon}e^{-iq_0T+i\mathbf q.\mathbf L} \nonumber\\
&\!\! \times \!\!& \int \dfrac{d^4\bar q}{(2\pi)^4}\dfrac{2\bar q_0}{\bar q^2- m_j^2+i \epsilon}e^{-i\bar q_0\bar T+i\mathbf{\bar q}.\mathbf{\bar L}}\nonumber\\
&\!\! \times \!\!& \int[d\mathbf p]f_z(\mathbf p,\mathbf P) \int d^4x e^{i(q+\bar q-p)x} \int d^4x_1 e^{-iqx_1} \int d^4x_2 e^{-i\bar qx_2}\nonumber\\
&\!\! \times \!\!& \int [d\mathbf k]f_{D_I}(\mathbf k,\mathbf K)e^{-ikx_1} \int [d\mathbf k']f^*_{D_F}(\mathbf k',\mathbf K')e^{ik'x_1} \int [d\mathbf k'']f^*_{l^-}(\mathbf k'',\mathbf K'')e^{ik''x_1}\nonumber\\
&\!\! \times \!\! & \int [d\mathbf {\bar k}]f_{\bar D_I}(\mathbf{\bar k},\mathbf{\bar K}) e^{-i\bar kx_2} \int [d\mathbf {\bar k'}]f^*_{\bar D_F}(\mathbf{\bar k'},\mathbf{\bar K'})e^{i\bar k'x_2}\int [d\mathbf {\bar k''}]f^*_{l^+}(\mathbf{\bar k''},\mathbf{\bar K''})e^{i\bar k''x_2}\nonumber\\
&\!\! \times \!\! & M_{j\bar D}(\bar k,\bar k', \bar k'')M_{jjZ}(p) M_{jD}(k,k',k'') \ .
\eea
Notice that the integration over $x$ in the recent equation leads to the $\delta$-Dirac function representing energy-momentum conservation in the source.
Hereafter, we assume, for simplicity, the momentum wave functions of the initial and final states to be Gaussian which are sharply peaked around the corresponding averaged momentum similar to
\be
f(\mathbf p,\mathbf p_i)=\left(\frac{\sqrt{2\pi}}{\sigma _p}\right)^{3/2} e^{\frac{-(\mathbf p-\mathbf p_i)^2}{4\sigma ^2_p}},
\label{G}
\ee
where $\sigma _p$, width of  momentum distribution, is assumed to be very smaller than the corresponding averaged momentum.
 Therefore, similar to the method presented in \cite{CC}, the amplitude of the total process can be written as
\bea\nonumber
\mathcal A_{\alpha\beta} &\!\! \propto\! \! & \dfrac{1}{\sqrt{3}}\sum _j U_{\alpha j}^*U_{\beta j}\int \dfrac{d^4q}{(2\pi)^4}\dfrac{2q_0}{q^2-m_j^2+i\epsilon}e^{-iq_0T+i\mathbf q.\mathbf L} \nonumber\\
&\!\! \times \!\! &\int \dfrac{d^4{\bar q}}{(2\pi)^4}\dfrac{2\bar{ q}_0}{\bar {q}^2-{m}_j^2+i {\epsilon}}e^{-i\bar q_0\bar T+i\mathbf{\bar q}.\mathbf{\bar L}}\int[d\mathbf p]f_z(\mathbf p,\mathbf P)(2\pi)^4\delta ^4(q+\bar q-p)\nonumber\\
&\!\! \times \!\! &\int d^4x_1  \exp\left[-i(q_0+E_{D_I}-E_{D_F}-E_{l^-})t_1+i(\mathbf q+\mathbf K-\mathbf K'-\mathbf K'')\mathbf x_1\right. \nonumber\\
&\!\! - \!\! & \left.\dfrac{(\mathbf x_1-\mathbf v_{D_I}t_1)^2}{4\sigma ^2_{xD_I}} - \dfrac{(\mathbf x_1-\mathbf v_{D_F}t_1)^2}{4\sigma ^2_{xD_F}}- \dfrac{(\mathbf x_1-\mathbf v_{l^-}t_1)^2}{4\sigma ^2_{xl^-}}\right]\nonumber\\
&\!\! \times \!\! &\int d^4x_2  \exp\left[-i(\bar{q}_0+\bar E_{{ D}_I}-\bar E_{{D}_F}-\bar E_{l^+})t_2+i(\mathbf{\bar q}+\mathbf{\bar K}-\mathbf{\bar  K'}-\mathbf{\bar  K''})\mathbf {x}_2\right. \nonumber\\
&\!\! - \!\! & \left.\dfrac{(\mathbf{ x}_2-\mathbf {v}_{\bar{D}_I}t_2)^2}{4\sigma ^2_{x\bar{D}_I}} - \dfrac{(\mathbf {x}_2-\mathbf {v}_{\bar{ D}_F}t_2)^2}{4\sigma ^2_{x\bar{ D}_F}}- \dfrac{(\mathbf {x}_2-\mathbf {v}_{l^+}t_2)^2}{4\sigma ^2_{xl^+}}\right]\nonumber\\
&\!\! \times \!\! & M_{j\bar D}(\bar k,\bar k', \bar k'')M_{jjZ}(p) M_{jD}(k,k',k'') \ ,
\label{AMP}
\eea
where $\sigma_x$'s are the position uncertainties which are related to the momentum ones through $\sigma _{x}\sigma _{p}\sim\dfrac{1}{2}$  and $\mathbf v$'s denote the group velocities of the corresponding particles.
Since the elements of matrix $M$ are smooth functions of the on-shell 4-momenta, whereas the wave packets of the external states are assumed to be sharply peaked at or near the corresponding mean momentum, one can replace $M$ by their values at the mean momenta and pull out of the integral.
Moreover, we define
$$
E_{D_I}-E_{D_F}-E_{l^-}=E_D,~~~~~~~~{\bar E}_{\bar D_I}-{\bar E}_{\bar D_F}-\bar E_{l^+}={\bar E_{\bar D}},
$$
$$
\mathbf K-\mathbf K'-\mathbf K''=\mathbf K_D,~~~~~~~~\mathbf {\bar K}-\mathbf {\bar K'}-\mathbf {\bar K''}=\mathbf {\bar K_{\bar D}},
$$
$$\mathbf v_D\equiv \sigma _{xD}^2(\dfrac{\mathbf v_{D_I}}{\sigma _{xD_I}^2}+\dfrac{\mathbf v_{D_F}}{\sigma _{xD_F}^2}+\dfrac{\mathbf v_{l^-}}{\sigma _{xl^-}^2}),~~~~~~~~\mathbf v_{\bar D}\equiv \sigma _{x{\bar D}}^2(\dfrac{\mathbf v_{{\bar D}_I}}{\sigma _{x{\bar D}_I}^2}+\dfrac{\mathbf v_{{\bar D}_F}}{\sigma _{x{\bar D}_F}^2}+\dfrac{\mathbf v_{l^+}}{\sigma _{xl^+}^2}),$$
$$\mathbf \Sigma _P\equiv \sigma _{xP}^2(\dfrac{\mathbf v^2_{P_I}}{\sigma _{xP_I}^2}+\dfrac{\mathbf v^2_{P_F}}{\sigma _{xP_F}^2}+\dfrac{\mathbf v^2_{l^-}}{\sigma _{xl^-}^2}),~~~~~~~~\mathbf \Sigma _{\bar D}\equiv \sigma _{x{\bar D}}^2(\dfrac{\mathbf v^2_{{\bar D}_I}}{\sigma _{x{\bar D}_I}^2}+\dfrac{\mathbf v^2_{{\bar D}_F}}{\sigma _{x{\bar D}_F}^2}+\dfrac{\mathbf v^2_{l^+}}{\sigma _{xl^+}^2}).$$
Therefore, using above issues and carrying out the integration over $x_1$ and $x_2$ one can write the amplitude as follows:
\bea\nonumber
\mathcal A_{\alpha\beta} &\!\! \propto\! \! & \dfrac{1}{\sqrt{3}}\sum _j U_{\alpha j}^*U_{\beta j}M_{j\bar D}(\bar K,\bar K', \bar K'')M_{jjZ}(P) M_{jD}(K,K',K'')(2\pi)^4 \nonumber\\
&\!\! \times \!\! &\int \dfrac{d^4q}{(2\pi)^4}\dfrac{2q_0}{q^2-m_j^2+i\epsilon}e^{-iq_0T+i\mathbf q.\mathbf L} \int \dfrac{d^4{\bar q}}{(2\pi)^4}\dfrac{2\bar{ q}_0}{\bar {q}^2- {m}_j^2+i {\epsilon}}e^{-i\bar q_0\bar T+i\mathbf{\bar q}.\mathbf{\bar L}}\nonumber\\
&\!\! \times \!\! &\int[d\mathbf p]f_z(\mathbf p,\mathbf P)\delta ^4(q+\bar q-p) e^{-S(q)} e^{-\bar S(\bar q)}\ .
\eea
where
\be
S(q)=\dfrac{(\mathbf {K}_D+\mathbf q)^2}{4\sigma ^2_{pD}}+\dfrac{[(q_0+E_D)-\mid\mathbf {K}_D+\mathbf q\mid\mathbf v_D]^2}{4\sigma ^2_{pD}\lambda_{D}},
\ee
and
\be
\bar S(\bar q)= \dfrac{(\mathbf {\bar K}_{\bar D}+\mathbf {\bar q})^2}{4\sigma ^2_{p\bar D}}+\dfrac{[(\bar{q}_0+E_{\bar D})-\mid\mathbf {\bar K}_{\bar D}+\mathbf {\bar q}\mid\mathbf v_{\bar D}]^2}{4\sigma ^2_{p{\bar D}}\lambda_{\bar D}},
\ee
with $\lambda _{D(\bar D)}\equiv\mathbf \Sigma _{D(\bar D)}-\mathbf v_{D(\bar D)}^2$.
Now, one should carry out the integration over the momentum of either propagating neutrino or propagating antineutrino. Here, we integrate over the momentum of antineutrino. After applying the following change in integration variable
$$
\mathbf p-\mathbf q=\mathbf p'\Longrightarrow \mathbf p=\mathbf p'+\mathbf q\Longrightarrow d^3\mathbf p=d^3\mathbf p',
$$
we have
\bea\nonumber
\mathcal A_{\alpha\beta} &\!\! \!\!\!\propto\!\! \!\!\! & \dfrac{1}{\sqrt{3}}\!\sum _j U_{\alpha j}^*U_{\beta j}M_{j\bar D}(\bar K,\bar K', \bar K'')M_{jjZ}(P) M_{jD}(K,K',K'')(2\pi)^4\nonumber\\
&\!\! \!\!\!\times \!\!\!\!\! &\int \!\!\dfrac{d^4q}{(2\pi)^4}\dfrac{2q_0}{q^2-m_j^2+i\epsilon}e^{-iq_0T+i\mathbf q.\mathbf L} e^{-S(q)}\!\!\int\!\![d\mathbf p']f_Z(\mathbf p^\prime+\mathbf q,\mathbf P)e^{-\bar S(p^\prime)}\nonumber\\
&&\times\dfrac{2E_{p'}}{p'^2-m_j^2+i\epsilon}e^{-i\bar E_j\bar T+i\mathbf p'.\mathbf {\bar{ L}}}\ ,
\eea
where $\bar E_j=E_p-q_0$, $p'^2=\Bar E_j^2-\mathbf {p'}^2$.
We perform the integral over $p'$ using the Grimus-Stockinger theorem \cite{16}
$$\int d^3\mathbf p^\prime\dfrac{\phi (\mathbf p')e^{i\mathbf p'.\mathbf L}}{{p'_a}^2-{{\mathbf p}^\prime}^2+i\epsilon}\longrightarrow ^{L\rightarrow \infty} -\dfrac{2\pi ^2}{ L}\phi (p'_a\hat{\mathbf L})e^{ip'_aL},$$
where $\hat{\mathbf L}=\frac{\mathbf L}{\mid\mathbf L\mid}$. This theorem is valid for a function $\phi$ which is differentiable at least three times such that $\phi$ itself and its first and second derivatives decrease at least as $\dfrac{1}{p'^2}$ as $\mid p'\mid \rightarrow \infty$. Performing the recent stage, one can write the amplitude as follows:
\bea\nonumber
\mathcal A_{\alpha\beta} &\!\! \!\!\propto\! \! \!\!&-\dfrac{2\pi ^2}{\bar L} \dfrac{1}{\sqrt{3}}\sum _j U_{\alpha j}^*U_{\beta j} M_{j\bar D}(\bar K,\bar K', \bar K'')M_{jjZ}(P) M_{jD}(K,K',K'')(2\pi)^4\nonumber\\
&\!\!\!\! \times \!\! \!\!&\int \dfrac{d^4q}{(2\pi)^4}\dfrac{4q_0(E_p-q_0)}{q^2-m_j^2+i\epsilon}e^{-iq_0T+i\mathbf q.\mathbf L} e^{-S(q)}f_Z(\bar{\mathbf p}_j+\mathbf q)e^{-\bar S(\bar p_j)}e^{-i\mid\bar{\mathbf p}_j\mid{\bar L} }e^{-iE_p\bar T}\,\nonumber\\
&&\times e^{iq_0\bar T},
\eea
in which $\mid \bar{\mathbf p}_j\mid=\sqrt{\bar E_j^2-{m_j}^2}$.
To carry out the integration over the neutrino 4-momentum, it will be more convenient for us to integrate first over $q^0$ and then over the components of $\mathbf q$. It is noticeable that since the pole at $q^0=-E_j+i\epsilon$ is not physical, the contribution to the integral is only given by the residue at the pole of the neutrino propagator at $q^0=E_j-i\epsilon$ . We obtain
\bea\nonumber
\mathcal A_{\alpha\beta} &\!\! \propto\! \! &\dfrac{4i\pi ^2}{\bar L} \dfrac{1}{\sqrt{3}}\sum _j U_{\alpha j}^*U_{\beta j} M_{j\bar D}(\bar K,\bar K', \bar K'')M_{jjZ}(P) M_{jD}(K,K',K'')(2\pi)^4\nonumber\\
&\!\! \times \!\! &e^{-iE_p\bar T}\int \dfrac{d^3q}{(2\pi)^3}f_Z(\bar{\mathbf p}_j+\mathbf q)(E_p-E_j(\mathbf q))e^{-iE_j(\mathbf q)(T-\bar T)}e^{i\mathbf q.\mathbf L} e^{-i\mid\bar{\mathbf p}_j\mid\bar L}\nonumber\\
&&\times  e^{-S(q)}e^{-\bar S(\bar p_j)}.
\eea
The remaining integration over $\mathbf q$ can be done by using saddle-point approximation at $\mathbf q=\mathbf p_j$. 
 We expand $E_j(\mathbf q)$ about $\mathbf q=\mathbf p_j$ as follows:
 \be
 E_j(\mathbf q)=E_j(\mathbf p_j)+(\mathbf q-\mathbf p_j)v_j+...,
 \ee
 where $v_j=\frac{\partial E_j}{\partial \mid\mathbf q\mid}$ at ${\mid\mathbf q\mid=\mid\mathbf p_j\mid}$. Also $S(\mathbf q)+\bar S(\mathbf p-\mathbf q)$ is expanded as follows;
 \be
 S(\mathbf q)+\bar S(\mathbf p-\mathbf q)=S(\mathbf p_j)+\bar S(\bar{\mathbf p}_j)+\frac 1 2 \frac{\partial^2(S+\bar S)}{\partial \mathbf q^2}(\mathbf q-\mathbf p_j)^2+...,
 \ee
 where the first derivative of $S+\bar S$ at $\mathbf q=\mathbf p_j$ vanishes and the second derivative is given by
\be
\dfrac{\partial ^2(S+\bar S)}{\partial q^2}=\dfrac{1}{2\sigma ^2_{pD}}+\dfrac{(\mathbf v_j-\mathbf v_D)^2}{2\sigma ^2_{pD}\lambda_D}+\dfrac{1}{2\sigma ^2_{p\bar D}}+\dfrac{(\mathbf v_j-\mathbf v_{\bar D})^2}{2\sigma ^2_{p\bar D}\lambda_{\bar D}}=\Omega _j.
\ee
Using the above issues, one can perform the integration over $d^3\mathbf q$. Consequently, the amplitude is obtained as follows:
\bea
\mathcal A_{\alpha\beta}&\!\!\!\! \propto\!\!\!\! &\frac{4i\pi ^2}{\bar L} \frac{1}{\sqrt{3}}\sum _j U_{\alpha j}^*U_{\beta j}M_{j\bar D}(\bar K,\bar K', \bar K'')M_{jjZ}(P) M_{jD}(K,K',K'')(2\pi)^4\nonumber\\
&&\,\,\,\,\,\,\,\,\,\,\times e^{-iE_p\bar T}(E_p-E_j)f_z(\mathbf P)e^{-i\mathbf P\bar L}\nonumber\\
&&\,\,\,\,\,\,\,\,\,\,\times\exp\left[-iE_j(T-\bar T)+i\mathbf p_j( L+{\bar L})-\dfrac{((L+{\bar L})-\mathbf v_j(T-\bar T))^2}{2\Omega _j})\right]\nonumber\\
&&\,\,\,\,\,\,\,\,\,\, \times \exp{(-S(p_j)-\bar S(\bar p_j))}.
\eea
The probability of the process is proportional to $\mid \mathcal A_{\alpha\beta}\mid ^2$. In a practical experimental setting $L$ and $\bar L$ are usually  fixed and known quantity while $T$ and $\bar T$ are not measured. Therefore, the probability of detecting a neutrino with flavor $\alpha$ and an antineutrino with flavor $\beta$ by the neutrino and antineutrino detectors located at the distances $L$ and $\bar L$ from the source, respectively, is obtained by the time average of $\mid \mathcal A_{\alpha\beta}\mid ^2$, which leads to
\bea
 P_{\alpha\beta}&\!\!\!\!\propto\!\!\!\!&\frac{1}{3}\sum _{j,k} U_{\alpha j}^*U_{\beta j}U_{\alpha k}U_{\beta k}^*N_jN^*_k\exp \Big\{i(\mathbf p_j-\mathbf p_k)( L+ {\bar L})-(S(p_j)+\bar S(\bar p_j))\nonumber\\
 &&\hspace{2cm}-(S(p_k)+\bar S(\bar p_k))
-\dfrac{i( L+{\bar L})(E_j-E_k)(\Omega _k\mathbf v_j+\Omega _j\mathbf v_k)}{(\Omega _k\mathbf v_j^2+\Omega _j\mathbf v_k^2)}\! \nonumber\\
&&\hspace{2cm}-\! \dfrac{(L+{\bar L})^2(\mathbf v_j-\mathbf v_k)^2}{2(\Omega _k\mathbf v_j^2+\Omega _j\mathbf v_k^2)}\! -\!\dfrac{(E_j-E_k)^2\Omega _j\Omega _k}{2(\Omega _k\mathbf v_j^2+\Omega _j\mathbf v_k^2)}\Big\} ,
\eea
where $$N_j=\dfrac{4i\pi ^2}{\bar L} M_{j\bar D}(\bar K,\bar K')M_{jjZ}(P) M_{jD}(K,K')(2\pi)^4e^{-iE_p\bar T}(E_p-E_j)e^{-ip_a\bar L}f_z(\mathbf P),$$ and $N^*_j$ is its complex conjugate. 
Since we are concerned with relativistic neutrinos, we use the following approximations.
 The differences between the energies and momenta of various mass eigenstates are due to the thin splitting of masses. Hence, we approximate
\be
E_j\simeq E+ \rho\dfrac{m_j^2}{2E},
\label{1q}
\ee
in which $E$ is the common neutrino energy when $m_i=0$ and $\rho$ is determined from the energy-momentum conservation \cite{qm}. Equation (\ref{1q}) leads to the following approximations
\be
p_j\simeq E+(\rho -1)\dfrac{m_j^2}{2E},
\label{2q}
\ee
and
\be
\mathbf v_j\simeq 1-\dfrac{m_j^2}{2E^2}.
\label{3q}
\ee
 Also, due to these approximations, one can easily show that
\be
\Omega _j\simeq 2\omega \sigma _x^2,
\ee
where $\sigma _x^2\equiv \sigma _{xD}^2+\sigma _{x{\bar D}}^2$ and
\be
\omega \equiv 1+\dfrac{\sigma _{xD}^2(1-\mathbf v_D)^2}{\sigma _x^2\lambda _D}+\dfrac{\sigma _{x\bar D}^2(1-\mathbf v_{\bar D})^2}{\sigma _x^2\lambda _{\bar D}}.
\ee
Moreover, one can see that the relativistic approximation leads $S+\bar S$ to be minimum. Therefore, in the relativistic approximation we obtain the following expression for the flavor-changing probability:
 \bea\label{qftresult}
 P_{\alpha\beta}\propto \dfrac{1}{3}\sum _{j,k} U_{\alpha j}^*U_{\beta j}U_{\alpha k}U_{\beta k}^*N_jN^*_k\exp\Big[&\!\!\!\!\!-&\!\!\!\!\!2\pi i\dfrac{L+\bar L}{L_{jk}^{osc}}-(\dfrac{L+\bar L}{L_{jk}^{coh}})^2\nonumber\\
 &\!\!\!\!\!-&\!\!\!\!\!2\pi ^2 \rho ^2\omega (\dfrac{\sigma _x}{L_{jk}^{osc}})^2\Big] ,
\eea
with the oscillation length $L_{jk}^{osc}$ and the coherence length $L_{jk}^{coh}$, for $j\neq k$, given by
\be
L_{jk}^{osc}\equiv \dfrac{4\pi E}{\Delta m_{jk}^2},~~~~~~~~~~L_{jk}^{coh}\equiv 2\sqrt{2\omega}\dfrac{2E^2}{\mid \Delta m_{jk}^2\mid}\sigma _x.
\ee
The exponent in the transition probability obtained for neutral current neutrino includes three terms; the first term leads to the usual oscillation pattern between the detectors, the second term indicates that the coherency condition is satisfied provided that the distance between the detectors is not larger than the coherence length and finally the third term shows that the position uncertainty due to the detection mechanisms must not be larger than the oscillation length. It is noticeable that
 \begin{itemize}
 \item the coherent propagation of both neutrino and antineutrino is not sufficient because the oscillation pattern is ceased if the distance between the detectors is larger than the coherence length. In fact, in quantum field theory approach, the conservation of energy-momentum due to the integration over the coordinates of the $Z_0$ decay vertex makes neutrino and antineutrino propagators entirely entangled.
\item the integration over the coordinates of the vertex of $Z_0$ decay gives energy-momentum conversation and the uncertainty of source is, practically, excluded from calculations. In other words, the source uncertainty does not play any role in the coherency of neutral current neutrinos and the detector uncertainties are analogues to the production and detection uncertainty in the case of he standard neutrino oscillation in the baseline $L+\bar L$.
 \end{itemize}

\section{Summary and Discussion}
 It has been shown that we can see neutrino oscillation pattern for $Z_0$ decay neutrinos provided that both neutrino and anti-neutrino are detected \cite{smirnov}. In this paper, we restudy this oscillation and corresponding decoherence issues via quantum field theory approach. We should emphasis that although, detection of two neutrinos is far from the experiment, the theoretical study of the neutral current neutrino oscillation leads to some nontrivial viewpoints about the theory of neutrino oscillation. In quantum field theory approach, neutrino and antineutrino are described by free propagators and the initial and final particle states are described by corresponding wave functions. The conservation of energy-momentum due to the integration over the coordinates of vertex of the $Z_0$ decay makes neutrino and antineutrino propagators entirely entangled. Therefore, the coherency of individual neutrino and antineutrino is not enough for oscillation, but the distance between the corresponding detectors have to be smaller than the coherence length. The other important result is related to the uncertainties of source and detectors; the source uncertainty does not play any role in the coherency of neutral current neutrinos and the detector uncertainties are analogues to the production and detection uncertainty in the case of he standard neutrino oscillation in the baseline $L+\bar L$.

{\bf Acknowledgement:}
The authors would like to thank Y. Farzan for her fruitful comment and S. M. Fazeli and R. Moazzemi for their useful discussions.

\end{document}